\documentclass[]{aastex631}

\usepackage{newtxtext,newtxmath}

\begin{document}

\title{Kink instability of flux ropes in partially-ionised plasmas}

\correspondingauthor{Giulia Murtas}
\email{giuliamurtas31994@gmail.com}

\author[0000-0002-7836-7078]{Giulia Murtas}
\affiliation{Los Alamos National Laboratory, Los Alamos NM, 87545, USA}
\affiliation{Department of Mathematics and Statistics, University of Exeter, Exeter EX4 4QF, UK}

\author[0000-0002-0851-5362]{Andrew Hillier}
\affiliation{Department of Mathematics and Statistics, University of Exeter, Exeter EX4 4QF, UK}

\author[0000-0002-4500-9805]{Ben Snow}
\affiliation{Department of Mathematics and Statistics, University of Exeter, Exeter EX4 4QF, UK}

\begin{abstract}

In the solar atmosphere, flux ropes are subject to current driven instabilities that are crucial in driving plasma eruptions, ejections and heating. A typical ideal magnetohydrodynamics (MHD) instability developing in flux ropes is the helical kink, which twists the flux rope axis. The growth of this instability can trigger magnetic reconnection, which can explain the formation of chromospheric jets and spicules, but its development has never been investigated in a partially-ionised plasma (PIP). Here we study the kink instability in PIP to understand how it develops in the solar chromosphere, where it is affected by charge-neutral interactions. Partial ionisation speeds up the onset of the non-linear phase of the instability, as the  plasma $\beta$ of the isolated plasma is smaller than the total plasma $\beta$ of the bulk. The distribution of the released magnetic energy changes in fully and partially-ionised plasmas, with a larger increase of internal energy associated to the PIP cases. The temperature in PIP increases faster also due to heating terms from the two-fluid dynamics. PIP effects trigger the kink instability on shorter time scales, which is reflected in a more explosive chromospheric flux rope dynamics. These results are crucial to understand the dynamics of small-scale chromospheric structures – mini-filament eruptions – that this far have been largely neglected but could significantly contribute to chromospheric heating and jet formation.

\end{abstract}

\keywords{Solar chromosphere (1479) --- Magnetohydrodynamics (1964) --- Solar magnetic reconnection (1504)}

\section{Introduction} \label{sec:intro}

In the solar chromosphere, the formation of highly dynamic features such as spicules \citep{1945ApJ...101..136R} and jets \citep{1951MNRAS.111..630M} is strongly bound to changes in connectivity of the magnetic field. The energy fuelling such explosive events can be released through the onset of magnetohydrodynamics (MHD) instabilities, that promote the conversion of stored magnetic energy through processes such as magnetic reconnection \citep{RevModPhys.82.603,2012ApJ...752L..12D}. The energy release associated with these processes promotes the formation of several types of dynamical structures, and can provide an important contribution to the chromospheric heating \citep{1945ApJ...101..136R,1951MNRAS.111..630M,1961ApJ...134..347O}.

Flux ropes, which are ubiquitous in the solar atmosphere, are twisted by convective motions, reaching an equilibrium that can be unstable to the kink modes \citep{1979SoPh...64..303H,1992PPCF...34..411H,2000mrmt.conf.....P}. Kink modes define a class of ideal, current-driven MHD instabilities that can become particularly important in triggering explosive events. When subjected to the kink instability, a plasma column bends and develops displacements that are transverse to the column axis and driven by an unbalance in magnetic pressure \citep{1992PPCF...34..411H}. The instability depends both on the length of the flux tube and the twist of magnetic field lines wrapped around it \citep{1990ApJ...350..428V}: if the field lines wrap multiple times along the length of the flux tube, then the system becomes unstable to kink modes \citep{2000mrp..book.....B}.

When a plasma column is traversed by an axial magnetic field, the $m = 1$ kink mode develops a helical deformation of the plasma column. Such deformation is magnified by the compression of the magnetic field lines at the internal side of the bends, with a consequent increase in magnetic pressure, and the expansion of field lines at the external side of the bends, that results in a decrease of magnetic pressure \citep{2004prma.book.....G}. The magnetic pressure imbalance leads to an increase in the distortion of the plasma column, which consequently results in a larger magnetic pressure gradient and a faster growth of the instability. The onset of the helical kink instability reduces the magnetic energy within the flux rope by converting the twist of the magnetic field lines in an helical deformation of the rope axis itself, namely the writhe \citep{2014PPCF...56f4012T}. Frequent observations of writhing in erupting filaments and prominences in the solar corona led to the suggestion, supported by many works, that the helical kink instability  could be responsible for triggering filament eruptions and CMEs \citep{1976PASJ...28..177S,2001ApJ...548..492S,2005ApJ...630L..97T,2005ApJ...630..543F}.

In systems where magnetic diffusivity is significant the kink instability develops in flux ropes, with an initial growth phase followed by a nonlinear resistive phase, characterised by current sheets formation and the onset of magnetic reconnection \citep{1988ApJ...330..474P,2009A&A...506..913H,2015RSPTA.37340266B,2018ApJ...863..172S}. The compression and expansion of magnetic field lines triggers the relaxation process \citep{PhysRevLett.33.1139, RevModPhys.58.741}, where reconnection straightens the magnetic field lines and lead to the release of the energy stored in the highly stressed and twisted field configuration. The magnetic field tends to relax toward a state of minimum magnetic energy defined by a force-free field, called the Taylor state \citep{PhysRevLett.33.1139}, where the magnetic helicity is preserved. This phase results in a less twisted configuration of the flux rope. During relaxation, magnetic energy is converted into kinetic energy and heat \citep{1984A&A...137...63H,2008A&A...485..837B}. The role of relaxation has been examined through several studies involving numerical simulations of cylindrical coronal loops \citep{2008A&A...485..837B, 2011A&A...525A..96B,2016A&A...589A.104G, 2016A&A...585A.159P}.

Flux ropes have been well studied in the fully-ionised coronal environment, where the kink instability has been identified in several observations \citep{2010ApJ...715..292S,2016ApJ...818..148L}, but it is unclear how these structures evolve in the lower atmosphere. Below the corona, twisted magnetic fields reconnect in regions where the plasma composition is heterogeneous and the plasma is only partially-ionised: this could be the case of micro-filament and mini-filament eruptions \citep{2015Natur.523..437S,2016ApJ...821..100S,2016ApJ...828L...9S,2019Sci...366..890S,2020ApJ...893L..45S}, taking place in small-scale loop systems observed in extreme ultraviolet and X-rays \citep{2021AAS...23821314H,2022arXiv220200370M, 2022arXiv220112314S}. The kink instability has been investigated as an explanation for jet formation through several numerical studies between the upper chromosphere \citep{Nishizuka_2008} and the lower corona \citep{1995Natur.375...42Y,2013ApJ...771...20M,2014ApJ...789L..19F}. However, numerical models that study unstable flux ropes in partially-ionised chromospheric plasma are still absent. Therefore, it is essential to investigate how the helical kink instability grows in a partially-ionised plasma, and the differences with the development of the same instability in a fully-ionised plasma.

The goal of this study is to examine the kink instability developing in single loops as a potential explanation for mini-filament eruptions in the solar chromosphere, and compare a fully-ionised plasma case (MHD) with two partially-ionised plasma cases (PIP). In Section \ref{methods}, the models of our simulations are presented. In Section \ref{results} results are discussed, with a focus on the energy release during the onset of magnetic reconnection within the nonlinear phase of the kink instability. Conclusions are presented in Section \ref{discussion}.

\section{Methods} \label{methods}

The simulations presented in this work are performed through the (P\underline{I}P) code \citep{2016A&A...591A.112H,2021A&A...645A..81S}. The equations are solved throughout the domain through a fourth-order central difference scheme, and the time updates are computed by a four-step Runge-Kutta scheme for time integration. 

The partially-ionised plasma is modelled through a two-fluid ion-neutral hydrogen plasma. Ions and neutrals are described by two separate sets of equations and coupled though elastic collisions, charge-exchange, collisional ionisation and recombination. The neutral fluid is governed by:
\begin{equation}
    \frac{\partial \rho_n}{\partial t} + \nabla \cdot (\rho_n \textbf{v}_n) = D,
\end{equation}
\begin{equation}
    \frac{\partial}{\partial t}(\rho_n \textbf{v}_n) + \nabla \cdot (\rho_n \textbf{v}_n \textbf{v}_n + p_n \textbf{I}) = \mathbf{R},
    \label{ch5:eq:force_neutral}
\end{equation}
\begin{equation}
    \frac{\partial e_n}{\partial t}  + \nabla \cdot [\textbf{v}_n (e_n + p_n)] = E,
    \label{ch5:eq:neutral_energy_2}
\end{equation}
\begin{equation}
    e_n = \frac{p_n}{\gamma -1} + \frac{1}{2} \rho_n v_{n}^{2},
    \label{ch5:neutral_energy}
\end{equation}
\begin{equation}
    T_n = \gamma \frac{p_n}{\rho_n},
    \label{ch5:eq:neutral_temperature}
\end{equation}
while the ionised fluid is governed by:
\begin{equation}
    \frac{\partial \rho_p}{\partial t} + \nabla \cdot (\rho_p \textbf{v}_p) = - D,
\end{equation}
\begin{equation}
    \frac{\partial}{\partial t}(\rho_p \textbf{v}_p) + \nabla \cdot \Bigg(\rho_p \textbf{v}_p \textbf{v}_p + p_p \textbf{I} - \textbf{B} \textbf{B} +  \frac{\textbf{B}^2}{2} \textbf{I} \Bigg) = - \mathbf{R},
    \label{ch5:eq:force_plasma}
\end{equation}
\begin{equation}
    \frac{\partial}{\partial t} \Bigg( e_p + \frac{B^2}{2}\Bigg) + \nabla \cdot [ \textbf{v}_p (e_p + p_p) -(\textbf{v}_p \times \textbf{B}) \times \textbf{B} \\
    + \eta (\nabla \times \textbf{B}) \times \textbf{B}] = -E - \Phi_I + A_{\text{heat}},
    \label{ch5:eq:3D_plasma_energy_equation}
\end{equation}
\begin{equation}
    \frac{\partial \textbf{B}}{\partial t} - \nabla \times (\textbf{v}_p \times \textbf{B} - \eta \nabla \times \textbf{B}) = 0,
\end{equation}
\begin{equation}
    e_p = \frac{p_p}{\gamma -1} + \frac{1}{2} \rho_p v_{p}^{2},
    \label{ch5:plasma_energy}
\end{equation}
\begin{equation}
    \nabla \cdot \textbf{B} = 0,
\end{equation}
\begin{equation}
    T_p = \gamma \frac{p_p}{2\rho_p}.
    \label{ch5:eq:plasma_temperature}
\end{equation}
The subscripts $p$ and $n$ identify physical quantities of the ion-electron plasma and of the neutral fluid respectively. The variables $\mathbf{v}$, $p$, $\rho$, $T$ and $e$ are the fluids velocity, gas pressure, density, temperature and internal energy, $\gamma= 5/3$ is the adiabatic index and \textbf{B} is the magnetic field. 

The terms $D$, $\mathbf{R}$ and $E$ are respectively the source terms for mass, momentum and energy transfer between the two species:
\begin{equation}
    D = \Gamma_{\text{rec}} \rho_p - \Gamma_{\text{ion}} \rho_n,
\end{equation}
\begin{equation}
    \mathbf{R} = - \alpha_c \rho_n \rho_p (\textbf{v}_n - \textbf{v}_p ) + \Gamma_{\text{rec}} \rho_p \mathbf{v}_p - \Gamma_{\text{ion}} \rho_n \mathbf{v}_n,
\end{equation}
\begin{equation}
    E = - \alpha_c \rho_n \rho_p \Bigg[ \frac{1}{2} (\textbf{v}_n ^2 -\textbf{v}_p ^2) + \frac{T_n - T_p}{\gamma (\gamma -1)} \Bigg] \\
    +\frac{1}{2} (\Gamma_{\text{rec}} \rho_p \mathbf{v}_p ^2 - \Gamma_{\text{ion}} \rho_n \mathbf{v}_n ^2) + \frac{\Gamma_{\text{rec}} \rho_p T_p - \Gamma_{\text{ion}} \rho_n T_n}{\gamma (\gamma -1)}.
\end{equation}
Both fluids are subject to the ideal gas law. The factor of 2 in equation~(\ref{ch5:eq:plasma_temperature}) is included to account for the electron pressure in the plasma fluid. 

The two-fluid collisional coupling is mediated by $\alpha_c (T_n , T_p ,v_D)$ \citep{1986MNRAS.220..133D,2018ApJ...869...23Z}, that is defined as:
\begin{equation}
\alpha_c = \alpha_c (0) \sqrt{\frac{T_n +T_p}{2}} \sqrt{1 + \frac{9\pi}{64} \frac{\gamma}{2(T_n + T_p)} v_D^2},
\label{ch5:alpha_c}
\end{equation}
where $\alpha_c (0)$ is the initial coupling and $v_D$ = $\mid$ \textbf{v$_n$} - \textbf{v$_p$} $\mid$ is the magnitude of the drift velocity between the neutral components and the hydrogen plasma. The collisional frequencies are determined by the product of $\alpha_c$ and the fluids density: the ion-neutral collisional frequency is defined as $\alpha_c \rho_n = \tau_{in}^{-1}$, and the neutral-ion collisional frequency is $\alpha_c \rho_p = \tau_{ni}^{-1}$, where $\tau_{in}$ and $\tau_{ni}$ are the ion-neutral and neutral-ion coupling time scales, respectively. While the coupling between ions and neutrals is governed by $\alpha_c$, the collisional coupling between ions and electrons is modelled by imposing a diffusivity $\eta$ in the system. The distribution for $\eta$ chosen in this work is spatially irregular and time-dependent: more details can be found in Section \ref{initial_conditions}. Initially, the two fluids are in thermal and ionisation equilibrium.

The terms $\Gamma_{\text{rec}}$ and $\Gamma_{\text{ion}}$ are the recombination and collisional ionisation rates for a hydrogen atom. The normalised empirical forms of the rates \citep{1997ADNDT..65....1V,2003poai.book.....S} are:
\begin{equation}
    \Gamma_{\text{rec}} = \frac{\rho_p}{\sqrt{T_p}} \frac{\sqrt{T_f}}{\xi_{p}(0)} \tau_{_{\text{IR}}},
    \label{ch5:eq:recombination_rate}
\end{equation}
\begin{equation}
    \Gamma_{\text{ion}} = \rho_p \Bigg( \frac{e^{- \chi} \chi^{0.39}}{0.232 + \chi} \Bigg) \frac{\hat{R}}{\xi_{p}(0)} \tau_{_{\text{IR}}},
    \label{ch5:eq:ionisation_rate}
\end{equation}
\begin{equation}
    \chi = 13.6 \frac{T_f}{T_{e0} T_p},
\end{equation}
\begin{equation}
    \hat{R} = \frac{2.91 \cdot 10^{-14}}{2.6 \cdot 10^{-19}} \sqrt{T_{e0}}.
    \label{ch5:eq:R_hat}
\end{equation}
Two characteristic temperatures appear in Equations (\ref{ch5:eq:recombination_rate})-(\ref{ch5:eq:R_hat}), and are based on a physical reference electron temperature $T_0$, expressed in Kelvin. $T_{e0}$ is the value of $T_0$ converted in electron volts. In our model we assume that protons and electrons are at the same temperature. Therefore, the choice of $T_0$ at the beginning of calculations determines the initial ion fraction $\xi_p (0)$. $T_f$ is a normalisation factor defined as
\begin{equation}
    T_f = \frac{\beta \gamma}{4} \frac{2 \xi_p (0)}{\xi_n (0) + 2 \xi_p (0)},
\end{equation}
and ensures that the reference temperature $T_0$ becomes the initial non-dimensional temperature of the simulation. The free parameter $\tau_{_{\text{IR}}}$ determines the ratio between recombination time scale and dynamic time: following our normalisation, $\tau_{_{\text{IR}}}$ states the initial recombination rate.

The terms $\Phi_I$ and $A_{\text{heat}}$ in equation~(\ref{ch5:eq:3D_plasma_energy_equation}) are associated to the ionisation potential and account for radiative losses. $\Phi_I$ approximates the energy removed by the system through ionisation, while $A_{\text{heat}}$ is an arbitrary heating term included to obtain an initial equilibrium. Their non-dimensional forms are:
\begin{equation}
    \Phi_I = \Gamma_{\text{ion}} \rho_n \hat{\Phi},
    \label{ch5:eq:Phi_I}
\end{equation}
\begin{equation}
    A_{\text{heat}} = \Gamma_{\text{ion}} (t = 0) \rho_n (t = 0) \hat{\Phi},
    \label{ch5:eq:A_heat}
\end{equation}
where $\hat{\Phi} = \frac{13.6 \beta}{2 K_B T_0}$ ensures consistency between the normalisation of ionisation potential and the equations modelling the system, which are normalised to the total Alfv\'en speed $v_A = 1$ and the total density $\rho_0$. As we include these heating/cooling terms associated to ionisation and recombination processes, the total energy of the system is not conserved in PIP. It is however conserved in fully-ionised MHD, as the solver is known to conserve energy in the absence of losses. More details on the atomic internal structure used to estimate the ionisation potential can be found in \cite{2021A&A...645A..81S}.

The physical variables in our system are non-dimensionalised as follows:

\begin{equation}
\begin{array}{lcl} \mathbf{r} \rightarrow r' \mathbf{\tilde{r}}, & \mathbf{B} \rightarrow B' \mathbf{\tilde{B}}, & \mathbf{v} \rightarrow v' \mathbf{\tilde{v}}, \\ t \rightarrow t' \tilde{t}, & \rho \rightarrow \rho' \tilde{\rho}, & P \rightarrow P ' \tilde{P}, \end{array}
\end{equation}

where $r'$ is the radius of the flux tube, $B' = 1$ is the initial axial field at $r = 0$, $v' = v_A = B'/ \sqrt{\rho'}$ as the permeability is set as $\mu_0 = 1$, the characteristic time scale $t' = t_A = r'/ v_A$ is the Alfv\'en transit time across the flux rope, $\rho' = 1$ is the total density and the total gas pressure $P' = \beta /2$. This non-dimensionalisation is consistent with the work in \cite{2009A&A...506..913H}.

\subsection{Initial conditions} \label{initial_conditions}

The initial setup is provided by a twisted magnetic flux tube in force-free equilibrium that is unstable to an ideal MHD kink instability. The initial non-potential, force-free magnetic field satisfies the condition:
\begin{equation}
    \nabla \times \mathbf{B} = \alpha(\mathbf{r}) \mathbf{B},
\end{equation}
where $\alpha = \mu_0 \mathbf{J} \cdot \mathbf{B} / B^2$ is the initial stress of the flux tube, and can relax through reconnection and reach a lower energy state, while preserving magnetic helicity. The field relaxes towards the minimum energy state, which must have a linear or constant $\alpha-$field profile: for this mechanism to work, the magnetic field must be sufficiently stressed beyond the relaxed state, before an instability releases the stored magnetic energy.

\begin{figure}
    \centering
    \includegraphics[width=\textwidth,clip=true,trim=0cm 5cm 0cm 0cm]{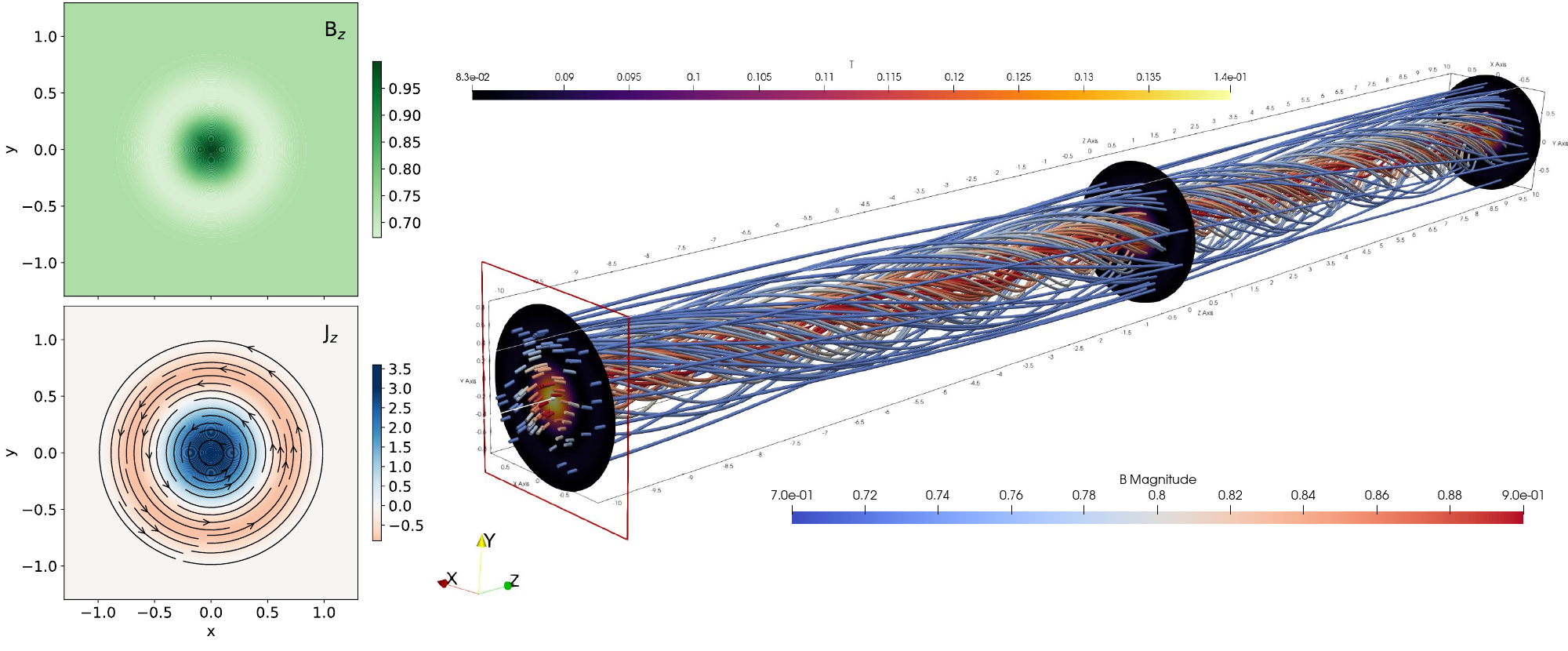}  
    \caption{Left: initial 2D profile of $B_z$ at $z = 0$ (top panel) and $J_z$ at $z = 0$ (bottom panel) in the $xy$-plane for all simulations. The black contour lines in the bottom panel represent the field lines of the $B_x B_y$ vector field. Right: initial 3D setup of the twisted magnetic field lines of case M1. The slices at $z = -10, 0, 10$ represent the magnitude of the plasma temperature.}
    \label{fig:1}
\end{figure}

In our simulations we model a cylindrical flux rope with a radius $r = 1$. The magnetic field components, consistent with the case studied in \cite{2009A&A...506..913H} for an initial smooth $\alpha$-profile, are reported in the following equations in cylindrical coordinates. For $r < 1$ we set:
\begin{equation}
    B_{\theta} (r) = \lambda r (1- r^2)^3 ,
    \label{ch5:eq:b_theta_lt_1}
\end{equation}
\begin{equation}
    B_z (r) = \sqrt{1- \frac{\lambda^2}{7} + \frac{\lambda^2}{7}(1 - r^2)^7 - \lambda^2 r^2 (1-r^2)^6} ,
    \label{ch5:eq:b_z_lt_1}
\end{equation}
\begin{equation}
    \alpha = \frac{2 \lambda (1 - r^2)^2 (1-4r^2)}{B_z (r)},
\end{equation}
while for $r \ge 1$ these components become:
\begin{equation}
    B_{\theta} = 0
    \label{ch5:eq:b_theta_gt_1},\,\,\,\,
    B_z = \sqrt{1 - \frac{\lambda^2}{7}},\,\,\,\,    \alpha = 0.
\end{equation}
This choice of the initial magnetic field components leads to a configuration with zero net-current. In the equations above, $\lambda$ is a constant parameter corresponding to a measure of the twist in the field. Due to the requirement from equation~(\ref{ch5:eq:b_z_lt_1}) that $B_z ^2$ must be positive, the value of $\lambda$ is constrained to be $\lambda < 64/965 \sqrt{1351} = 2.438$. In this work we impose $\lambda = 1.8$, consistent with \cite{2009A&A...506..913H}: the choice of these initial conditions also leads to a change of sign for $\alpha$ at $r = 0.5$. 
Figure \ref{fig:1} shows the initial profiles of $B_z$ and the current density component $J_z$ in the $xy$-plane ($z=0$) set for all simulations. The current density profile is continuous across the domain, and a smooth transition occurs from $J \ne 0$ to $J = 0$ at $r = 1$. The three-dimensional form of the magnetic field can be observed in Figure \ref{fig:1}, compared to the initial plasma temperature of case M1.

The plasma $\beta$ associated with the bulk pressure is 0.1, consistent with the upper chromosphere \citep{2001SoPh..203...71G}. Chromospheric neutrals play an important role in the decay of the non force-free components of the photospheric field, as it transitions to the force-free coronal field \citep{2000ApJ...533..501G,2004A&A...422.1073K, 2009ApJ...705.1183A}. In a two-fluid system, the balance with the Lorentz force can only be achieved by varying the plasma pressure: as the charges have an effective $\beta$ of 0.02, as such we expect the system to be approximately force-free. Therefore, a force-free equilibrium is a suitable starting point for this work.

%\begin{figure*}
%    \centering
%    \includegraphics[width=0.9\textwidth,clip=true,trim=0cm 0cm 0cm 0cm]{2.pdf}  
%    \caption{Initial 3D setup of the magnetic field lines of the flux rope in case M1. The slices at $z = -10, 0, 10$ represent the magnitude of the plasma temperature.}
%    \label{fig:2}
%\end{figure*}

The calculations are run with a non-uniform, current-dependent resistivity, that accounts for electron-ion collisions and is determined as follows:
\begin{equation}
    \eta = \eta_0 + \begin{cases} 0, & |J| < J_{\text{crit}} \\ \eta_1, & |J| > J_{\text{crit}} \end{cases}
\end{equation}
where $\eta_0$ is the background uniform resistivity, set equal to $\eta_0=10^{-4}$, and $\eta_1 = 10^{-3}$ is the anomalous resistivity component that is only switched on when the magnitude of the current exceeds a critical value, chosen to be $J_{\text{crit}} = 5$ to be above the initial maximum current magnitude determined by the initial conditions. This choice of diffusivity, resolved by the grid, is similar to the work of \cite{1999ApJ...517..990A}, also used by \cite{2008A&A...485..837B} and \cite{2009A&A...506..913H}.
As ambipolar diffusion represents strongly coupled two-fluid effects in a single-fluid regime \citep{2024mpsp.book..203S}, its role is already covered by the more complex two-fluid physics of our setup. As such, no separate resistive term (i.e. ambipolar diffusion, \citealp{1994ApJ...427L..91B,1995ApJ...448..734B,2021RAA....21..232C}) is included to account for the effects of neutral-charge interactions.
%\textcolor{red}{Neutral-charge interactions increase the resistivity in magnetized plasmas (i.e. \citealp{2021RAA....21..232C}). Ambipolar diffusion is traditionally used in single-fluid models to represent this effect \citep{1994ApJ...427L..91B,1995ApJ...448..734B}, and can be derived from two-fluid modelling \citep{2024mpsp.book..203S}. As the effects of ambipolar diffusion is already accounted for by our two-fluid model, we do not include a separate resistive term.} %However, ions and neutrals tend to decouple at the low chromospheric densities  \citep{2017PPCF...59a4038K}. This requires a two-fluid approach \citep{1989ApJ...340..550Z,2011PhPl...18k1210S,2015PASJ...67...96S,2011PhPl...18k1211Z,2012ApJ...760..109L,2013PhPl...20f1202L,2015ApJ...799...79N}, where the coupling terms describe how ion-neutral collisions vary as functions of the speed, density and temperature of the plasma components.

In order to trigger the instability in the system, a small velocity perturbation is included. The initial velocity perturbation, imposed in both the ion and neutral fluids to break the initial equilibrium, is a $xy$-plane white noise of the form:
\begin{equation}
    v_{x,p}, v_{y,p} = 0.05 e^{-(x^2 + y^2)} \cdot (\mbox{random noise}).
\end{equation}
The perturbation results in a larger velocity magnitude at the centre of the flux rope, perpendicular to the flux rope axis, and a smaller velocity magnitude at the extremities of the flux rope.  As this perturbation is set to be the same for all cases, its effects might vary following the different coupling between ions and neutrals, resulting in the onset of different modes. Evidence of this process is discussed in Section \ref{results}.

The simulations are run in a Cartesian computational domain $x = [-2, 2]$, $y = [-2, 2]$ and $z = [-10, 10]$ with a uniform grid. Although the analytical equilibrium is presented in cylindrical coordinates, the field components have been mapped onto the Cartesian grid to provide the initial states for the simulations. The domain is resolved by $496 \times 496 \times 868$ grid points, corresponding to a cell size of $\Delta x = 8 \cdot 10^{-3}$, $\Delta y = 8 \cdot 10^{-3}$ and $\Delta z = 2.3 \cdot 10^{-2}$. All boundaries are periodic, mimicking previous studies of instabilities in flux ropes \citep{shafranov1957structure,1958PhFl....1..421K}. Following this choice, the system simulates a potentially infinite number of infinitely long flux ropes, but the boundaries are set sufficiently far that the ropes never interact with each other, and the kink instability can be studied for a single structure. 

\section{Results} \label{results}

We analyse the kink instability in three simulations, corresponding to a fully-ionised plasma case (M1) and two PIP cases (P1 and P2), run at different collisional coupling. M1 represents the limit for a completely coupled system ($\alpha_c \rightarrow \infty$), where neutrals and charges act as a single fluid. P1 corresponds to a weakly coupled ion-neutral plasma, where collisions occur on timescales $\sim \rho^{-1}$ for a reference density $\rho$, while P2 is run with a stronger collisional coupling between the fluids, with collisions occurring on timescales $\sim 0.1 \rho^{-1}$. The parameter $\tau_{\text{IR}}$ is varied in the PIP cases to maintain the same proportionality between collision frequencies and ionisation/recombination rates. The full array of parameters is shown in Table~\ref{table:1}.
\begin{deluxetable*}{c c c c c c c c}
\tabletypesize{\small}
\tablewidth{0pt} 
\tablecaption{List of the simulation parameters. \label{table:1}}
\tablehead{
\colhead{ID} & \colhead{Model}& \colhead{$\alpha_c (0)$} & \colhead{$\xi_p$ (0)} & \colhead{$T_0$ (K)} & \colhead{$\tau_{\text{IR}}$} & \colhead{$\tau_{in} (0)$}& \colhead{$\tau_{ni} (0)$}
} 
\startdata 
 M1 & MHD & $\infty$ & 1 & - & - & - & - \\
 P1 & PIP & 1 & $10^{-1}$ & $1.28 \cdot 10^{4}$ & $10^{-5}$ & 1.11 & 10 \\
 P2 & PIP & $10$ & $10^{-1}$ & $1.28 \cdot 10^{4}$ & $10^{-4}$ & 0.11 & 0.1 \\
\hline \hline
\enddata
\end{deluxetable*}

\begin{figure*}
    \centering
    \includegraphics[width=0.8\textwidth,clip=true,trim=2cm 0cm 0.5cm 0cm]{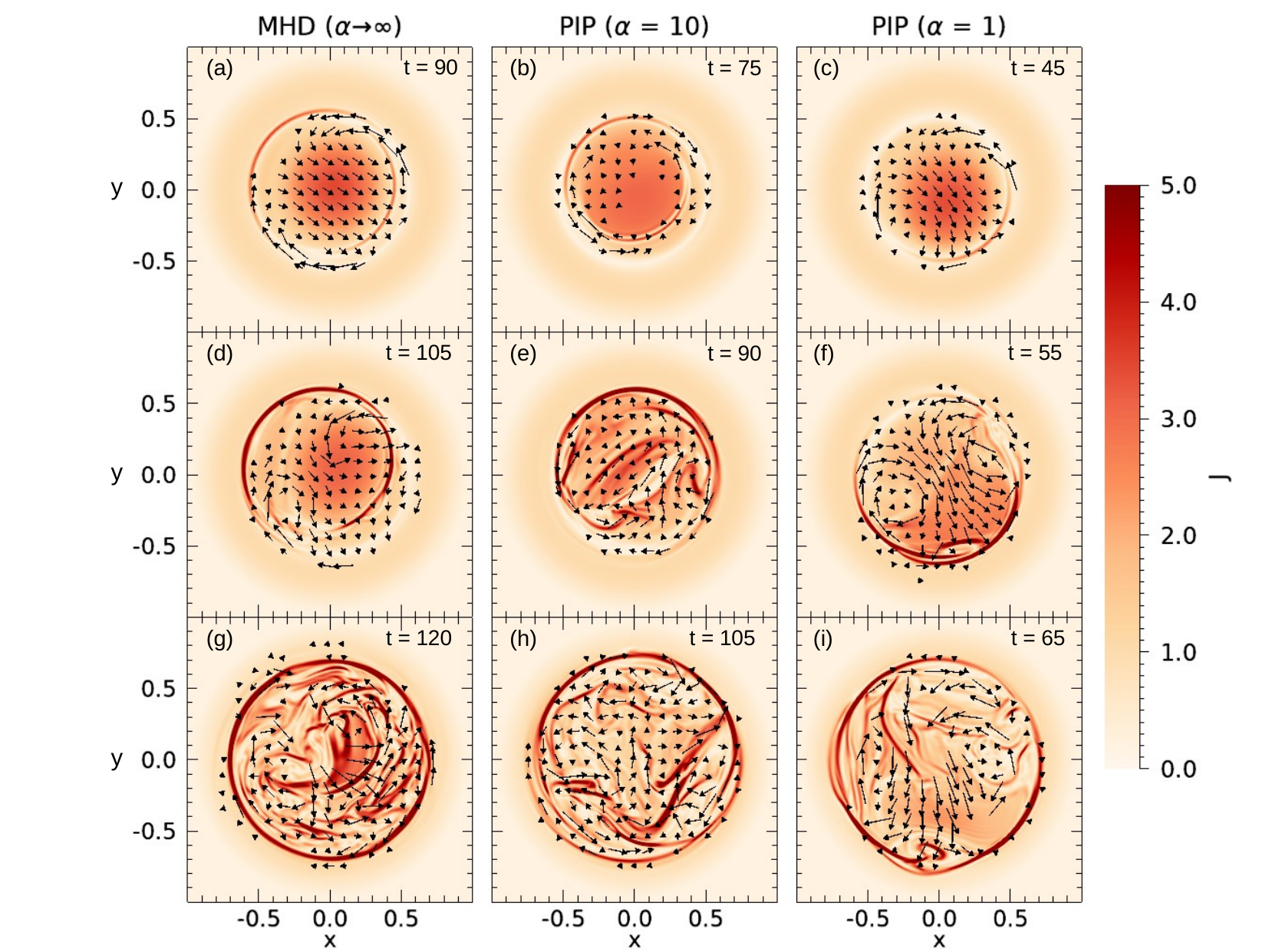}  
    \caption{Contour plot of the current density magnitude $J$ at $z$ = 0, showing the development of the kink instability in M1 (left), P2 (centre) and P1 (right). Each row displays the same stage of development of the kink instability. Panels ($a$), ($b$) and ($c$) show the onset of the kink instability. Panels ($d$), ($e$) and ($f$) display the evolution at a later stage, and the formation on current sheets inside the flux rope. Panels ($g$), ($h$) and ($i$) show the flux rope expanding and the formation of further current sheets.  The arrows show the direction of the plasma flow $\mathbf{v}_p$. Times are given in the same non-dimensional unit for all three cases. $J$ is saturated by fixing the maximum value of the colour scale to the critical value $J_{\text{crit}} = 5$, in order to enhance the structures characterised by strong currents.}
    \label{fig:3}
\end{figure*}

The development of the kink instability is shown in Figure \ref{fig:3} for three evolution stages in both fully-ionised plasmas (M1) and partially-ionised plasmas (P1 and P2).
At the onset of the instability, the current density is larger at the centre of the flux rope, as shown by panels ($a$), ($b$) and ($c$). Under the distortion generated by the kink, the plasma flow pushes the field lines to form current sheets, that appear in panels ($d$), ($e$) and ($f$) of Figure \ref{fig:3} as dark, thin and elongated regions where $J > 5$. As the magnetic tension reduces with the straightening of the field lines, and the gas pressure increases with the thermal energy released from reconnection, the flux rope expands. During the expansion phase, shown at early times in panels ($d$), ($e$) and ($f$), current sheets form at different locations and the reconnection process moves to regions further away from the centre of the flux rope. At later stages, shown in panels ($g$), ($h$) and ($i$) reconnection occurs mostly near the external boundary of the flux rope for P1 and P2, while in case M1 several structures of high concentrations of current density are still present around the centre of the flux rope.

As shown by Figure \ref{fig:3}, where the same instability phase appears in the bottom panels at $t = 120$ for M1, $t = 105$ for P2 and $t = 65$ for P1, respectively, the kink instability evolves much faster in partially-ionised plasmas than in fully-ionised plasmas. By comparing the two PIP cases, the flux rope expansion is faster when the collisional coupling $\alpha_c$ is smaller. The same development stage of the instability occurs earlier ($t = 65$) for case P1 (where $\alpha_c (0) = 1$) than case P2 ($t = 105$, where $\alpha_c (0)$ is set to be equal to 10). This occurs as the growth of the kink instability scales with the Alfv\'en speed, which is inversely proportional to the density. In case M1 the fluid is completely ionised, and the ion density coincides with the bulk density. In both PIP cases, run with the same bulk density as M1, the effective Alfv\'en speed is larger as it depends on a combination of the smaller ion density and the density of the neutrals that are collisionally coupled to the charges: for this reason, the kink instability grows faster. The lower coupling in case P1 speeds up the reconnection rate and pushes the flux rope to expand faster than in case P2, where the stronger coupling between ions and neutrals results in the evolution of the system more closely resembling that of a single, completely ionised fluid.

\begin{figure*}
    \centering
    \includegraphics[width=0.9\textwidth,clip=true,trim=0cm 15cm 0cm 0.2cm]{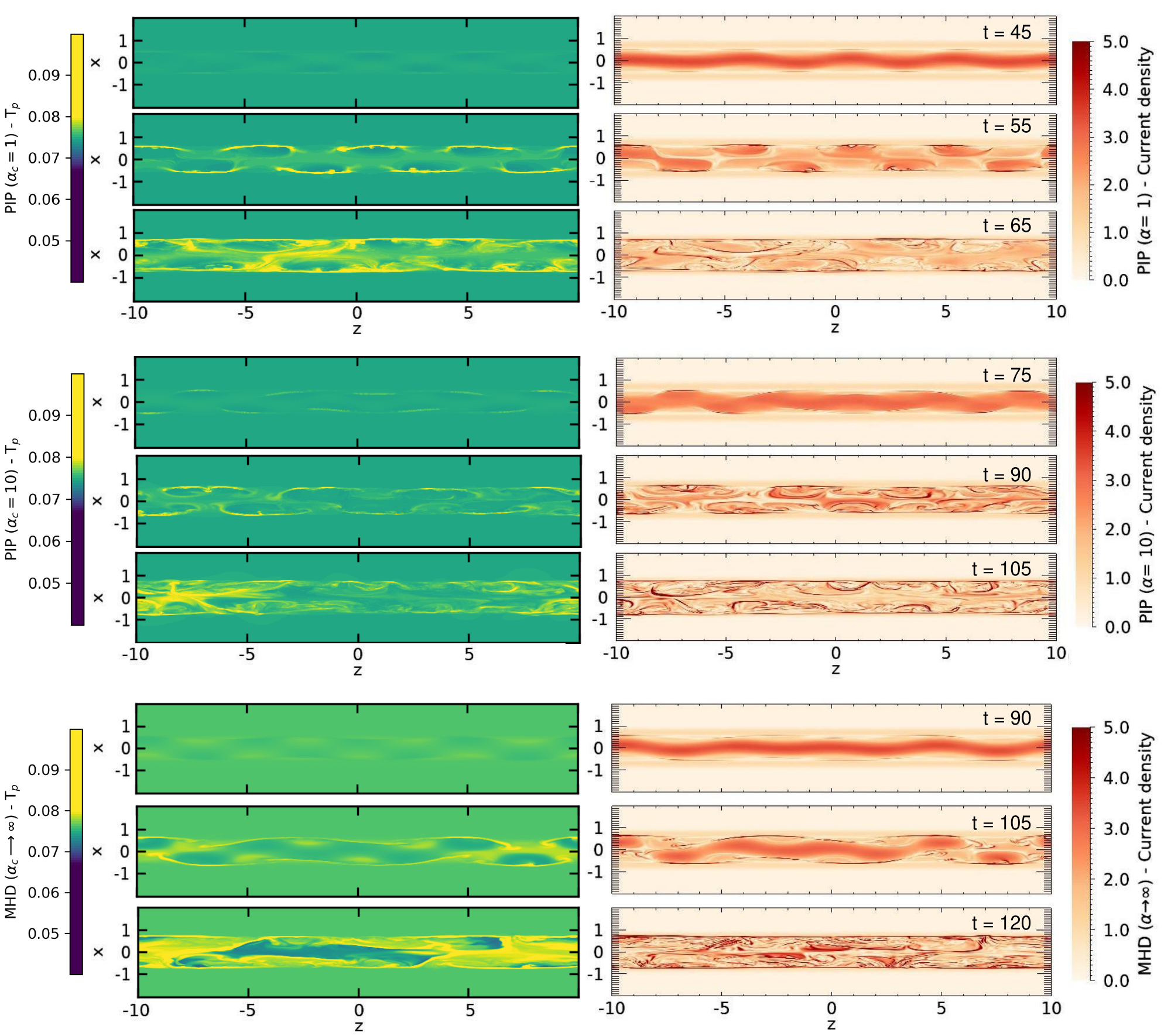}  
     \caption{Contour plot of the plasma temperature $T_p$ (left) and current density magnitude $J$ (right) in the $xz-$plane ($y = 0$), showing the development of the kink instability in case P1 (top panels), P2 (central panels) and M1 (bottom panels). Each set of panels displays the same stage of development of the kink instability shown in Figure \ref{fig:3}. $T_p$ is saturated by fixing the maximum value of the colour scale to the value 0.1, to enhance the plasma structures where the largest temperature increase occurs. Similarly, $J$ is saturated by fixing the maximum value of the colour scale to the critical value $J_{\text{crit}} = 5$ in order to enhance the structures characterised by strong currents.}
    \label{fig:4}
\end{figure*}
The charges temperature $T_p$ and the current density magnitude at the three instability phases in Figure \ref{fig:3} are shown in Figure \ref{fig:4} in the $xz-$plane ($y = 0$) laying along the flux rope axis for cases P1 (top six panels), P2 (central six panels) and M1 (bottom six panels). The writhe of the flux rope axis is observed in the top two panels of each set, corresponding to $t_{\text{P}_1} = 45$, $t_{\text{P}_2} = 75$ and $t_{\text{M}_1} = 90$. In case P1, where the two fluids are almost decoupled, four crests located at regular distances from each other are observed along the $z-$axis, which suggests the onset of a single, dominant mode of the instability. Similarly, four crests are obtained in case M1. In the intermediate case, however, it is possible to identify five crests, with a new, central peak appearing around $z = 0$. While a dominant mode is observed both in the PIP case with the lowest coupling and in the full coupling limit (M1), the very presence of oscillations with crests located at irregular distances in P2 suggests the overlapping of two modes of the instability. Therefore, the partially coupled neutrals leads to the selection of different modes of the kink instability.

\begin{figure*}
    \centering
    \includegraphics[width=0.8\textwidth,clip=true,trim=1cm 8.5cm 2cm 9cm]{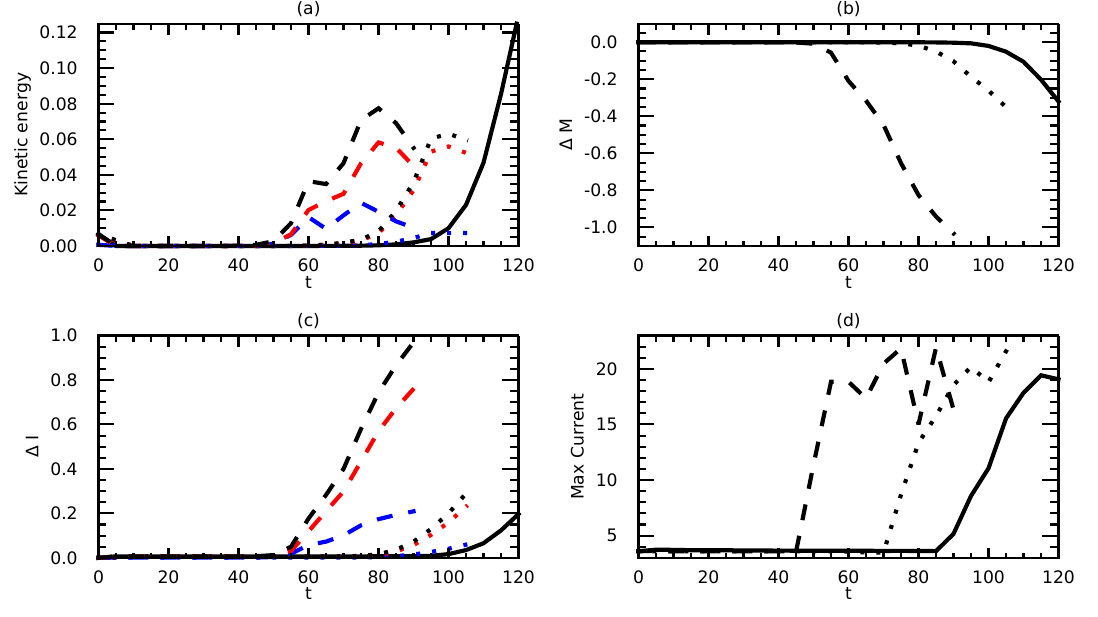} 
    \caption{Kinetic energy ($a$), time variation of ($b$) magnetic energy $\Delta M$ and ($c$) internal energy $\Delta I$, and  maximum current density ($d$) for M1, P1 and P2. In all four panels, the solid lines are associated to variables of the MHD case (M1), the dashed lines refer to case P1 and the dotted lines refer to case P2. In panels ($a$) and ($c$) blue lines are associated to the plasma energy components, red lines are associated to the neutral energy components and black lines are associated to the bulk energy component (only plasma for case M1, plasma and neutrals for cases P1 and P2). The kinetic energy is the only energy component that is perturbed initially.}
    \label{fig:5}
\end{figure*}
Figure \ref{fig:5} shows the time variation of the energy components integrated over the volume $V$ of the simulation box, and the maximum current density of cases M1, P1 and P2. In order to better show the variation of the energy terms, the net variation of internal and magnetic energy are displayed. The total kinetic energy is shown fully, and it differs from zero at $t = 0$ due to the initial velocity perturbation.

The passage from the linear growth phase to the nonlinear reconnection phase of the kink instability is identified by the decrease of magnetic energy $\Delta M$ (panel $b$ of Figure \ref{fig:5}), defined as:
\begin{equation}
    \Delta M = \int_V \frac{B^2}{2} dV - \int_V \frac{B^2 (t = 0)}{2} dV,
\end{equation}
and the increase of kinetic and internal energy components (panels $a$ and $c$ of Figure \ref{fig:5}) due to the energy conversion from reconnection, which for case M1 starts at $t \sim 90$, for case P1 is at $t \sim 50$ and for case P2 is at $t \sim 80$. The maximum current density reached in all three cases is approximately similar, as suggested by the trend of $J$ displayed in panel ($d$).

Regarding the distribution of the converted magnetic energy, the individual contributions of plasma (blue) and neutrals (red) to the kinetic energy in panel ($a$) and the total kinetic energy, defined as:
\begin{equation}
    K_{_{\text{TOT}}} = \int_V \Bigg( \frac{1}{2} \rho_p \mathbf{v}_p^2 + \frac{1}{2} \rho_n \mathbf{v}_n^2 \Bigg) dV,
\end{equation}
are smaller than the MHD case, as shown by the dashed black line (P1) and the dotted black line (P2), while the variation of the total internal energy $I_{_{\text{TOT}}}$, shown in panel ($c$) and calculated as:
\begin{equation}
    \Delta I_{_{\text{TOT}}} = \int_V \Bigg( \frac{p_p + p_n}{\gamma - 1} - \frac{p_p (t = 0) + p_n (t = 0)}{\gamma - 1} \Bigg) dV,
\end{equation}
is larger in the PIP cases than in the MHD case (solid black line) when compared at the same growth phase of the instability, despite smaller single contribution from charges and neutrals.

The difference in the increase of the bulk internal energy between PIP and MHD cases means that the temperature of cases P1 and P2 grows faster. In general, it must be expected that the temperature is higher in the PIP cases than in the MHD case: this comes from the inclusion of further heating terms that result from the interaction of charges and neutrals. This will be discussed in more detail in Section \ref{heating}.

Following the work of \cite{2008A&A...485..837B}, the instability growth rate can be calculated as:
\begin{equation}
    \sigma = \frac{1}{2} \frac{d\log(KE)}{dt}.
\end{equation}
Therefore, in this work, the linear growth rate of the kink instability is estimated as the angular coefficient of the slope given by the initial increase of the natural logarithm of the total kinetic energy (black lines in panel $a$ Figure \ref{fig:5}), divided by a factor of 2. The dimensionless growth rate of the three cases is, respectively,
\begin{itemize}
    \item $\sigma$ (P1) $ = 0.18 \pm 0.02$, in the interval $t = 20-60$,
    \item $\sigma$ (P2) $ = 0.118 \pm 0.006$, in the interval $t = 30-75$,
    \item $\sigma$ (M1) $ = 0.082 \pm 0.004$, in the interval $t = 55-115$,
\end{itemize}
where the error is determined as the standard deviation from the linear fit of the kinetic energy logarithm.

The estimated growth rate decreases at the increase of $\alpha_c$. The variation in growth rate between the fully coupled case (M1) and the least collisionally coupled case (P1) is of a factor of 2.2. The larger growth rate comes as a consequence of the variation in effective plasma density: in case of a fully decoupled case ($\alpha = 0$) run at the PIP ion fraction $= 0.1$, the growth of the instability is expected to be a factor $\sim \sqrt{10}$ $(\sim 3.16)$ larger than the MHD case presented here. The value of the growth rate for our intermediate coupling cases prove that there is a direct response of the coupling in affecting the instability growth rate.

\subsection{Two-fluid effects}
\label{two_fluid}

\begin{figure*}
    \centering
    \includegraphics[width=0.95\textwidth,clip=true,trim=0cm 0cm 0cm 0cm]{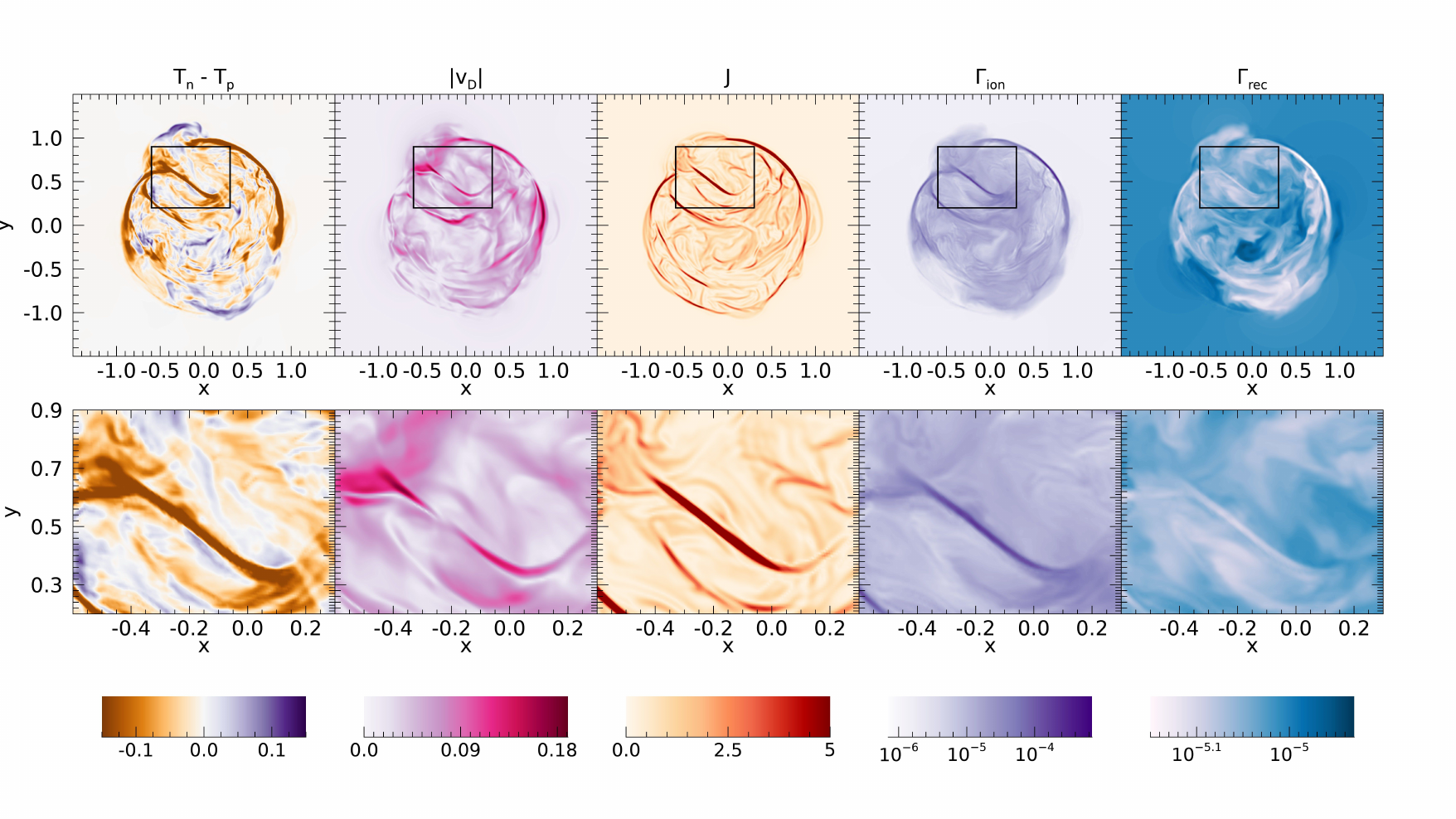}
    \caption{Contour plot of ($a$) the temperature difference $T_n - T_p$, ($b$) the drift velocity magnitude $|v_D|$, ($c$) the current density magnitude $J$, ($d$) the ionisation rate $\Gamma_{\text{ion}}$ and  ($e$) the recombination rate $\Gamma_{\text{rec}}$ for the PIP case P1 at $t = 90$ (top panels), and details of the same variables in the area around a current sheet (bottom panels). The variables are shown at the centre of the flux rope ($z = 0$). The magnitude of ionisation and recombination rates is presented with a logarithmic color scale. The area covered by the bottom panels corresponds to the region identified in the black boxes in the top panels.}
    \label{fig:6}
\end{figure*}

In a partially-ionised plasma, the interaction of ions and neutrals can lead to further dynamics that might affect the physics of the kink instability. Here case P1 is examined at $t = 90$, a late stage of the nonlinear evolution of the kink instability. Figure \ref{fig:6} shows the temperature difference $T_n - T_p$, the magnitude of the drift velocity between fluids $v_D = |\mathbf{v}_n - \mathbf{v}_p|$, the ionisation rate $\Gamma_{\text{ion}}$, the recombination rate $\Gamma_{\text{rec}}$ and the current density magnitude $J$ at the centre of the flux rope ($z = 0$). The whole section of the flux rope is shown in the top panels, while the detail of a smaller area surrounding a current sheet is displayed in the bottom panels.

The temperature distribution differs between neutral and plasma, as shown by the panels in column ($a$) of Figure \ref{fig:6} where golden areas represent the regions where the plasma is hotter than the neutrals and purple areas identify the regions where the neutral fluid is hotter than the plasma. The plasma temperature exceeds the neutral temperature mostly in regions of high current density magnitude, as shown by the comparison between the panels in column ($a$) and ($c$), while patches of higher neutral temperature than the plasma are observed outside the reconnecting regions. The maximum temperature difference reached in these areas ($\sim 0.17$) is very large when compared to the background temperature of 0.075 of both species, and dimensionally is equivalent to a difference of about $3 \cdot 10^4$ K. The magnitude of the drift velocity is also very high, reaching values of about a fifth of the bulk Alfv\'en speed, mostly in correspondence of high current density structures as shown by the comparison with $J$. This means that in high current density regions the two fluids are mostly decoupled, a feature that is also shown by the large difference in temperature in these areas, with peaks in plasma temperature within the current sheets and peaks in neutral temperature immediately outside. 

In P1, ionisation and recombination processes occur with a comparable increase inside the flux rope. Initially $\Gamma_{\text{ion}} = 10^{-6}$ and $\Gamma_{\text{rec}} = 10^{-5}$, values that remain constant in the background plasma. Spikes of ionisation rate are observed within the areas of larger current density and $T_p$, with a maximum value $\Gamma_{\text{ion}} = 2.4 \cdot 10^{-3}$ at $t = 90$: inside the current sheet, the plasma compression and high temperature allow to ionise the neutrals dragged by collisions with the ions. The larger ionisation rate leads to a thickening of the current sheet, as noted by \cite{2022PhPl...29f2302M}. Outside the longer current sheets, recombination rates increase proportionally to the plasma temperature, as shown by the comparison with the temperature difference map in column ($a$), with peaks of $\Gamma_{\text{rec}} = 1.2 \cdot 10^{-5}$ at $t = 90$. A similar trend is found for the case with a larger ion-neutral coupling (P2). In case P2, the background $\Gamma_{\text{ion}} = 10^{-5}$ and $\Gamma_{\text{rec}} = 10^{-4}$, with peaks of $\Gamma_{\text{ion}} = 4.4 \cdot 10^{-2}$ and $\Gamma_{\text{rec}} = 1.4 \cdot 10^{-4}$ at $t = 105$.

As shown by the zoom-in of the area around a current sheet, displayed in the bottom panels of Figure \ref{fig:6}, peaks of drift velocity (column \textit{b}) occur in the outflow regions of the current sheet, while in the inflow both neutral and plasma velocity are comparable. This has previously been observed in the results presented in \cite{2021PhPl...28c2901M} and \cite{2022PhPl...29f2302M}, where an evident decoupling of the species occurred in the outflow regions of current sheets. The expulsion of neutrals from the current sheet lead to a more turbulent mixing in the regions around, and to local peaks of the neutral temperature.

\begin{figure*}
    \centering
    \includegraphics[width=0.7\textwidth,clip=true,trim=0cm 0cm 0cm 0cm]{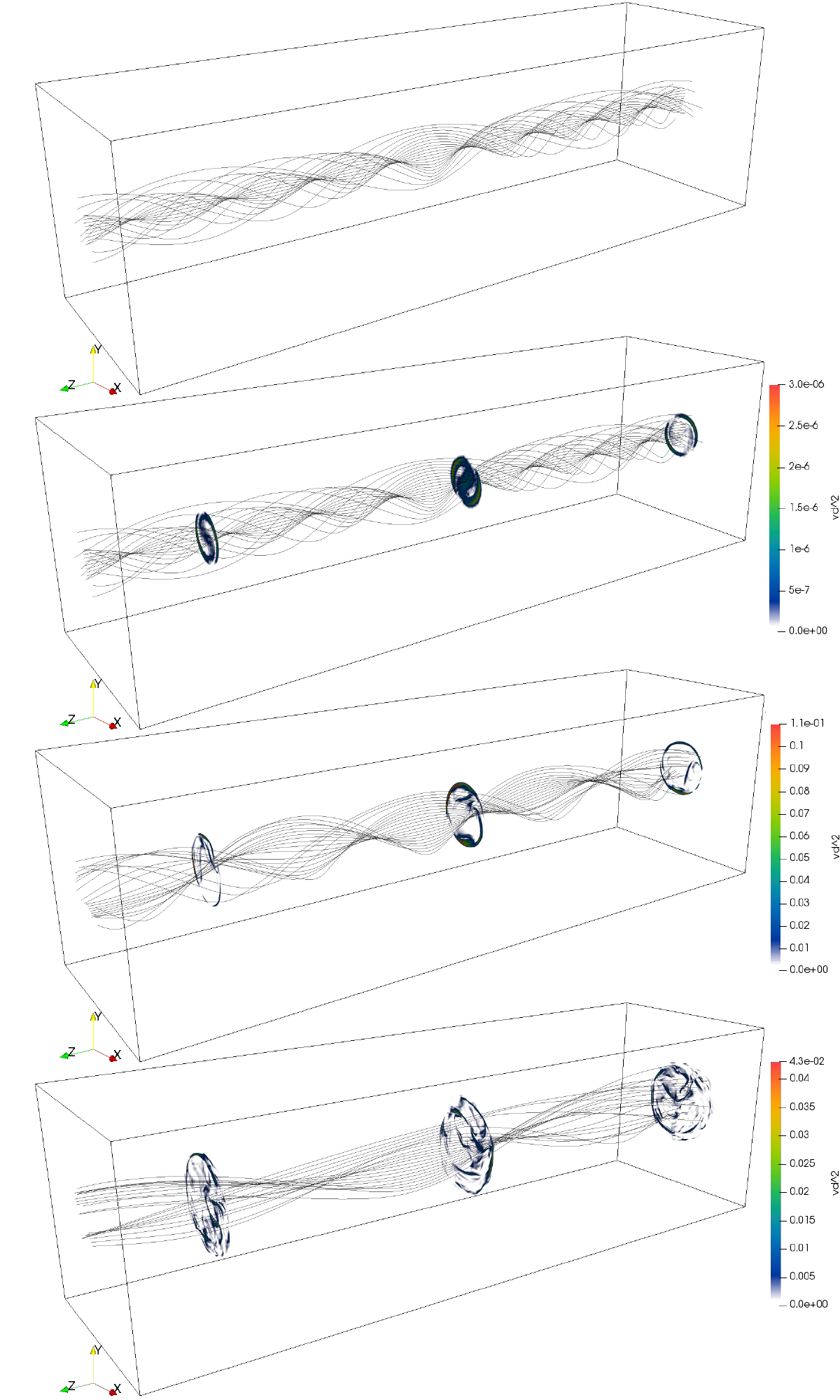}  
    \caption{Time series of the kink instability of case P1 showing magnetic field lines (grey) and slices of the drift velocity squared $(\mathbf{v}_n - \mathbf{v}_p)^2$. The times showed are, from top to bottom, $t = 0$, $t = 30$, $t = 60$ and $t = 90$.}
    \label{fig:7}
\end{figure*}
Figure \ref{fig:7} shows the time evolution of the kink instability and the variation of the drift velocity at three locations of the flux rope. At all times the drift velocity produces similar structures of comparable magnitude along the flux rope. In the early stages of the simulation ($t = 0$) characterised by the largest writhe, the drift velocity is negligible. The first structures in $v_D$ appear when the nonlinear phase of the instability begins ($t = 30$), and their distribution along a spiral structures follows the relaxation of the magnetic field lines. At $t = 60$ the drift velocity increases are concentrated primarily at the external surface of the flux rope, where the first current sheets are formed, as expected from the distribution of $v_D$ magnitude shown in Figure \ref{fig:6} and discussed above. Finally, the velocity structures created by the drift increase in complexity with the increasing turbulent motions occurring within the flux rope at later times ($t = 90$).

\subsection{Temperature and heating terms} \label{heating}

Beyond the term of Ohmic heating, the high velocity drifts observed in partially-ionised plasmas and discussed in Section \ref{two_fluid} can lead to heating generated by collisions taking place between ions and neutrals at these high speeds. In this Section we examine the evolution of the fluids temperature and of the heating terms with the growth of the kink instability.
\begin{figure}
    \centering
    \includegraphics[width=0.6\textwidth,clip=true,trim=0cm 0cm 0cm 0cm]{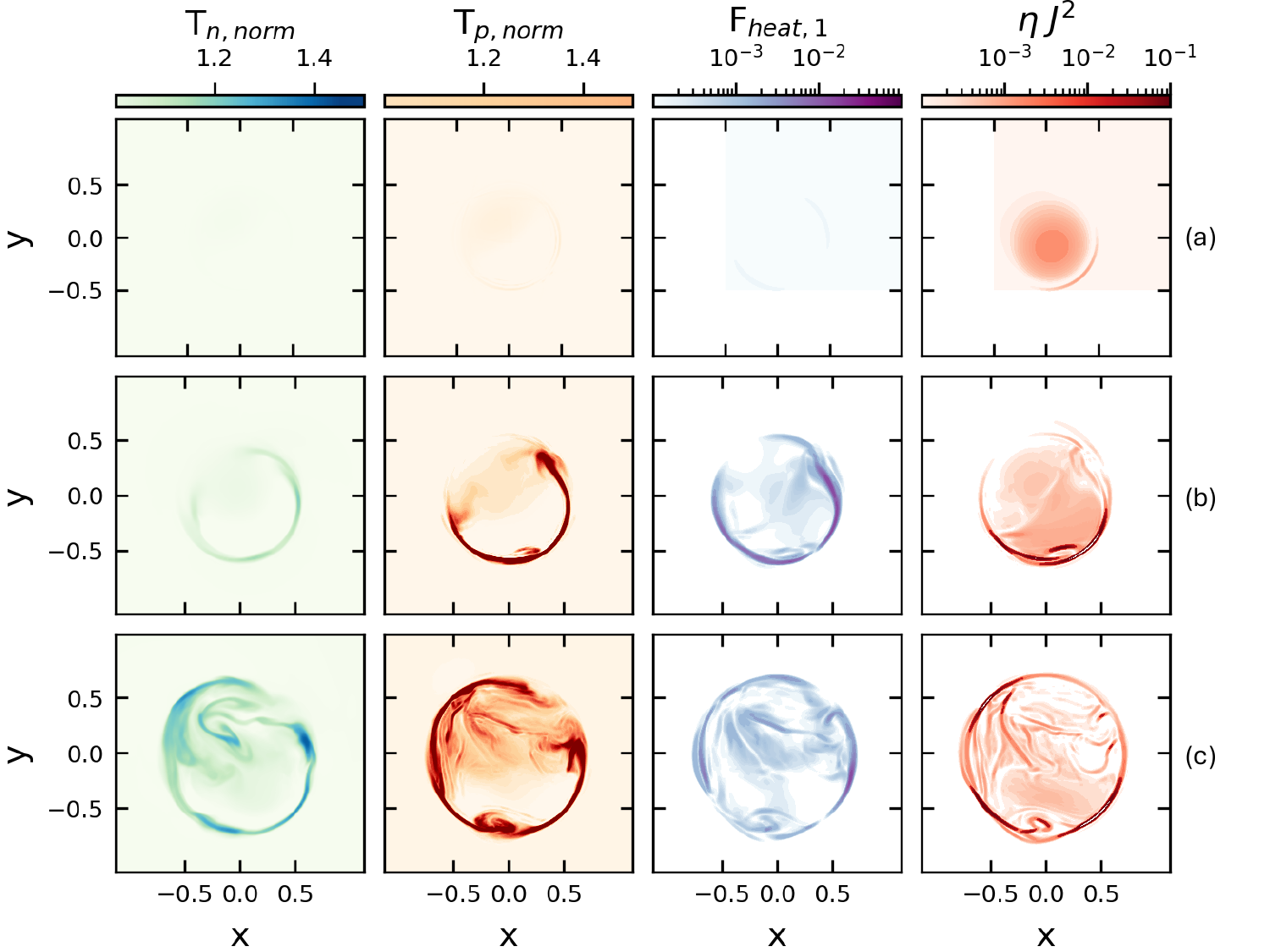}
    \caption{Left to right: neutral temperature $T_{n,\text{norm}}$ and plasma temperature $T_{p,\text{norm}}$ normalized by the respective background temperatures, collisional frictional heating $F_{\text{heat},1}$ and Ohmic heating $\eta J^2$ of case P1 at $t = 45$ (row $a$), $t = 55$ (row $b$) and $t = 65$ (row $c$).}
    \label{fig:8}
\end{figure}
\begin{figure}
    \centering
    \includegraphics[width=0.6\textwidth,clip=true,trim=0cm 0cm 0cm 0cm]{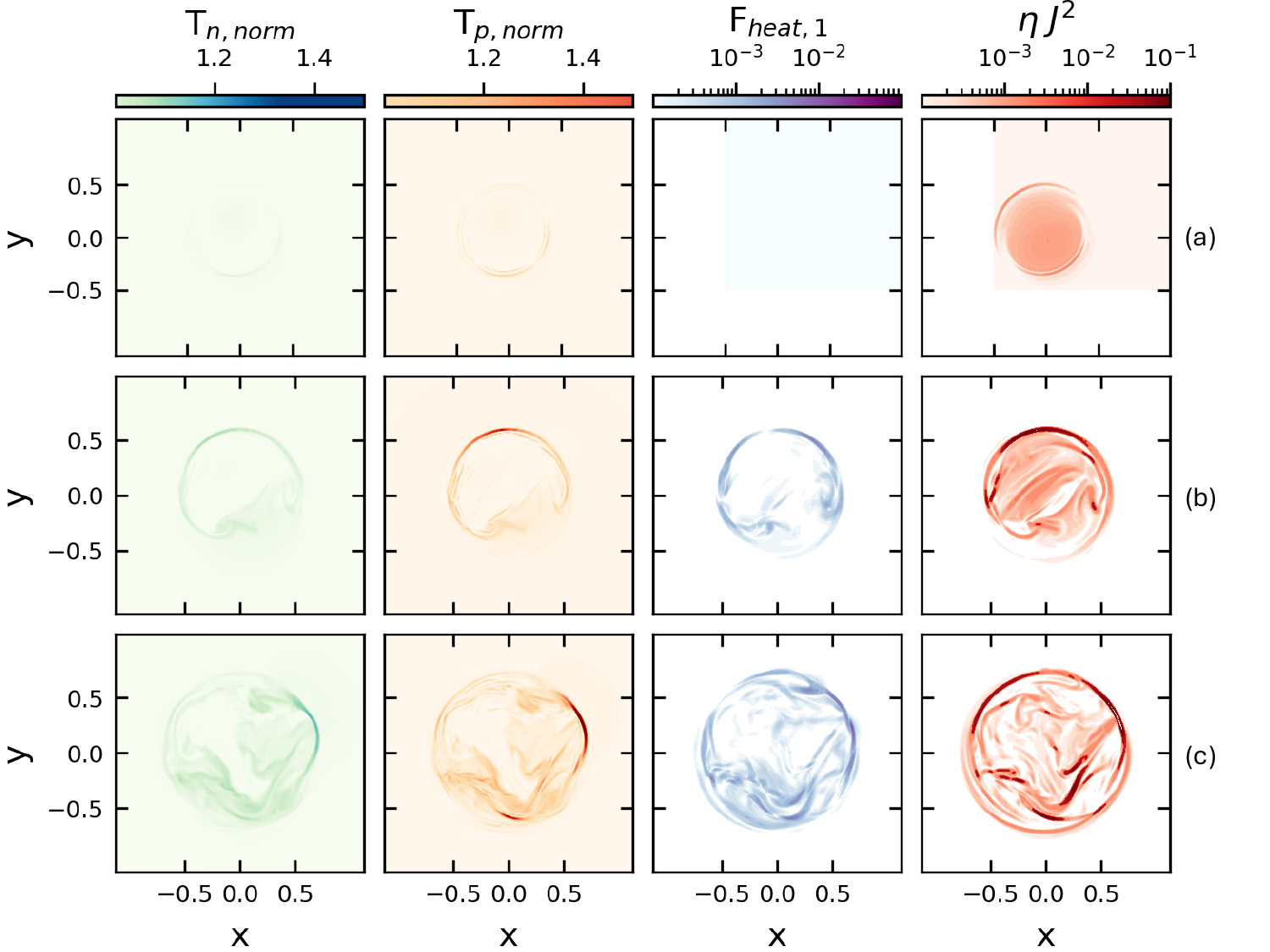}
    \caption{Left to right: neutral temperature $T_{n,\text{norm}}$ and plasma temperature $T_{p,\text{norm}}$ normalized by the respective background temperatures, collisional frictional heating $F_{\text{heat},1}$ and Ohmic heating $\eta J^2$ of case P2 at $t = 75$ (row $a$), $t = 90$ (row $b$) and $t = 105$ (row $c$).}
    \label{fig:9}
\end{figure}

The time variation of both neutral and plasma temperatures are shown in Figure \ref{fig:8} for case P1 and Figure \ref{fig:9} for case P2 at the same evolution phases displayed in Figure \ref{fig:3}. In a partially-ionised plasma, fluids can be heated through Ohmic heating $\eta J^2$, generated by the current dissipation in regions at high current densities, and by frictional heating, which is generated by the drift between ions and neutrals. Frictional heating is characterised by two different components \citep{2023MNRAS.525.4717S}. The first component is a collisional term related to the drift velocity between ions and neutrals. The non-dimensional form of the collisional frictional heating is:
\begin{equation}
    F_{\text{heat},1} = \frac{1}{2} \alpha_c (T_n, T_p, v_D) \rho_n \rho_p v_D ^2,
    \label{ch5:eq:3D_frictional_heating}
\end{equation}
and depends on both the collisional frequencies and the drift velocity between charges and neutrals. The second term adding to frictional heating is linked to the work done through ionisation-recombination processes, and takes the form:
\begin{equation}
    F_{\text{heat},2} = \Gamma_{\text{rec}} \rho_p \mathbf{v}_p^2 - (\Gamma_{\text{rec}} \rho_p + \Gamma_{\text{ion}} \rho_n) \mathbf{v}_n \cdot \mathbf{v}_p + \Gamma_{\text{ion}} \rho_n \mathbf{v}_n^2.
    \label{ch5:eq:ionrec_frictional_heating}
\end{equation}

The contour maps of the Ohmic heating and the collisional frictional heating terms are shown in Figure \ref{fig:8} for case P1 and \ref{fig:9} for case P2. In both Figures, the colour scales are set equal for an easier comparison between the two PIP cases. At the beginning of the kink instability both $T_p$ and $T_n$ of cases P1 and P2 present very small variations with respect to the initial conditions, as shown by the panels in row ($a$) of Figure \ref{fig:8}-\ref{fig:9}. The plasma temperature starts increasing when the first current sheets form inside the flux rope, as shown in rows ($b$) and ($c$) of Figure \ref{fig:8}-\ref{fig:9}, and as a consequence of the thermal coupling between the two fluids the neutral temperature starts to increase in the same areas of high $T_p$. The Ohmic term is larger in magnitude than frictional heating, as shown by the comparison of the heating terms magnitude in both Figure \ref{fig:8}-\ref{fig:9}, and its action is localised to the current sheets structures. Outside these areas, collisional frictional heating provides a comparable contribution to heating both charges and neutrals. The frictional heating produced by ionisation-recombination processes is a few orders of magnitude smaller in both PIP cases, reaching a peak of $1.78 \cdot 10^{-5}$ for P1 at $t = 65$ and $4.13 \cdot 10^{-5}$ for P2 at $t = 105$, hence its contribution can be considered negligible.

The temperatures and heating terms of the two PIP cases can be compared at similar growth stages of the instability, and across the central section of the flux rope. In case P1, where the initial collisional coupling is $\alpha_c (0) = 1$, the plasma temperature at the later stages displays the largest increase, reaching peaks of $T_{p,P1} \sim 0.25$, 3.3 times the background temperature in the reconnecting regions (corresponding to peaks of $\sim 4.2 \cdot 10^4$ K), against $T_{p,P2} \sim 0.21$ for case P2 where $\alpha_c (0) = 10$, which is about 2.7 times the background temperature (corresponding to $\sim 3.4 \cdot 10^4$ K). From the underlying empirical ionisation and recombination equations \ref{ch5:eq:recombination_rate}-\ref{ch5:eq:ionisation_rate}, the equilibrium ionisation fraction can be defined as a function of temperature only \citep{2021A&A...645A..81S}, i.e., 
\begin{gather} \label{eqn:ioneq}
\xi_p=\frac{1}{\frac{F(T)}{G(T)}+1},
\end{gather}
where $F(T),G(T)$ are the temperature dependent components of the recombination and ionisation rates respectively. The temperature maxima in both P1 and P2 cases is high enough that the plasma should be, locally, completely ionised ($\xi_p \approx 1$) based on Equation \ref{eqn:ioneq}. However, Equation \ref{eqn:ioneq} is based on steady-state equilibrium, whereas the simulation is time-dependent and allows departures from equilibrium to exist. The ionisation and recombination rates are relatively slow, with $\Gamma_{\rm ion} \approx 10^{-3}$ in the reconnection region, as shown in Figure \ref{fig:6} for case P1 where $\Gamma_{\rm ion}$ peaks at $2.4 \cdot 10^{-3}$ at $t = 90$. As such, the timescale for ionisation-recombination equilibrium to be obtained is long compared to the life of the feature. Within the reconnection region, the ionisation fraction is roughly 0.1 and as such, the local neutral density is far larger than would be predicted by ionisation-recombination equilibrium. The larger temperature increase in P1 (the case with the smallest ion-neutral coupling) depends on the difference in the effective plasma $\beta$, which is smaller in the most decoupled case as a lower coupling results in a lower gas density. The largest increases in neutral temperatures for the two PIP cases are closer, being $T_{n,P1} \sim 0.11$ and $T_{n,P2} \sim 0.13$, 1.5 and 1.74 times the background temperature respectively, despite the difference in the magnitude of the plasma temperature.

A second difference is observed in the magnitude of the heating components between case P1 and P2. While the Ohmic heating $\eta J^2$ is similar in both cases, the maximum being $\sim 0.25$ for case P1 and $\sim 0.26$ for case P2 in the last phase of the instability (row $c$ of Figure \ref{fig:8}-\ref{fig:9}), the contribution of the collisional frictional heating is larger in case P2, being $\sim 0.04$ against a maximum value of $\sim 0.014$ for case P1. The difference in frictional heating comes from the difference in the collision coupling between the two cases. Collisional frictional heating is directly proportional to $\alpha_c$ as shown by the definition in equation~(\ref{ch5:eq:3D_frictional_heating}), and directly affects the increase in temperature of both charges and neutrals.

\begin{figure}
    \centering
    \includegraphics[width=0.6\columnwidth,clip=true,trim=0cm 0cm 0cm 0cm]{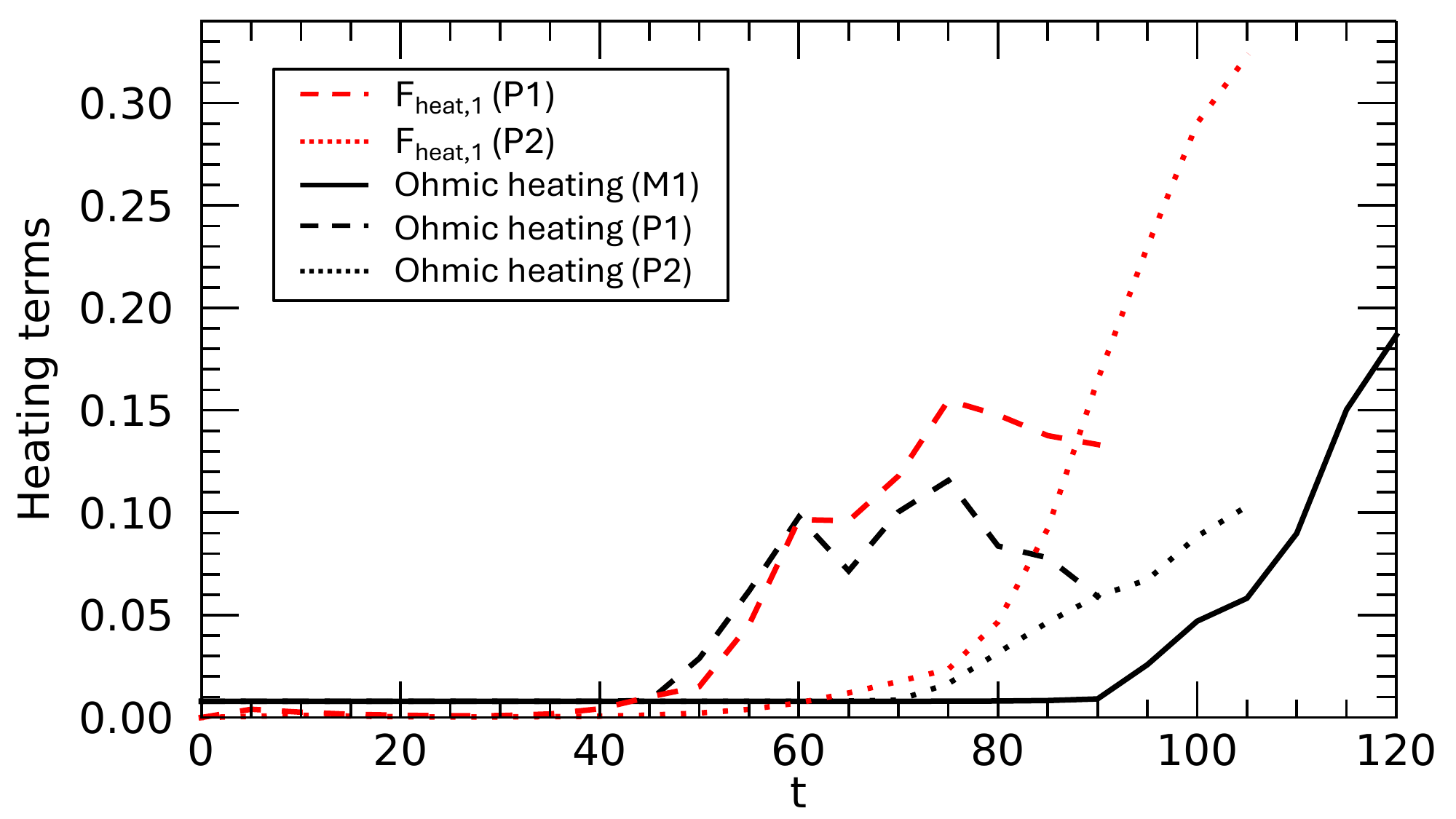}
    \caption{Global Ohmic heating (black) and collisional frictional heating (red) as a function of time for case M1 (solid line), P1 (dashed line) and P2 (dotted line).}
    \label{fig:10}
\end{figure}
While the collisional frictional heating is smaller locally than the Ohmic heating, its total contribution is larger than Ohmic heating in both PIP cases. This is shown in Figure \ref{fig:10}, where Ohmic heating and collisional frictional heating integrated over the domain are displayed as a function of time. This occurs because the Ohmic heating, despite being larger in magnitude, is localised in the few current sheet structures at higher current densities, while the frictional heating produced by the drift between charges and neutrals is produced more uniformly inside the whole flux rope. As already noticed from the magnitude of the collisional frictional heating in Figure \ref{fig:8}-\ref{fig:9}, P2 develops larger amounts of collisional frictional heating than P1 due to the stronger coupling between fluids.

\begin{figure}
    \centering
    \includegraphics[width=0.6\textwidth,clip=true,trim=1.7cm 0.5cm 1.7cm 0cm]{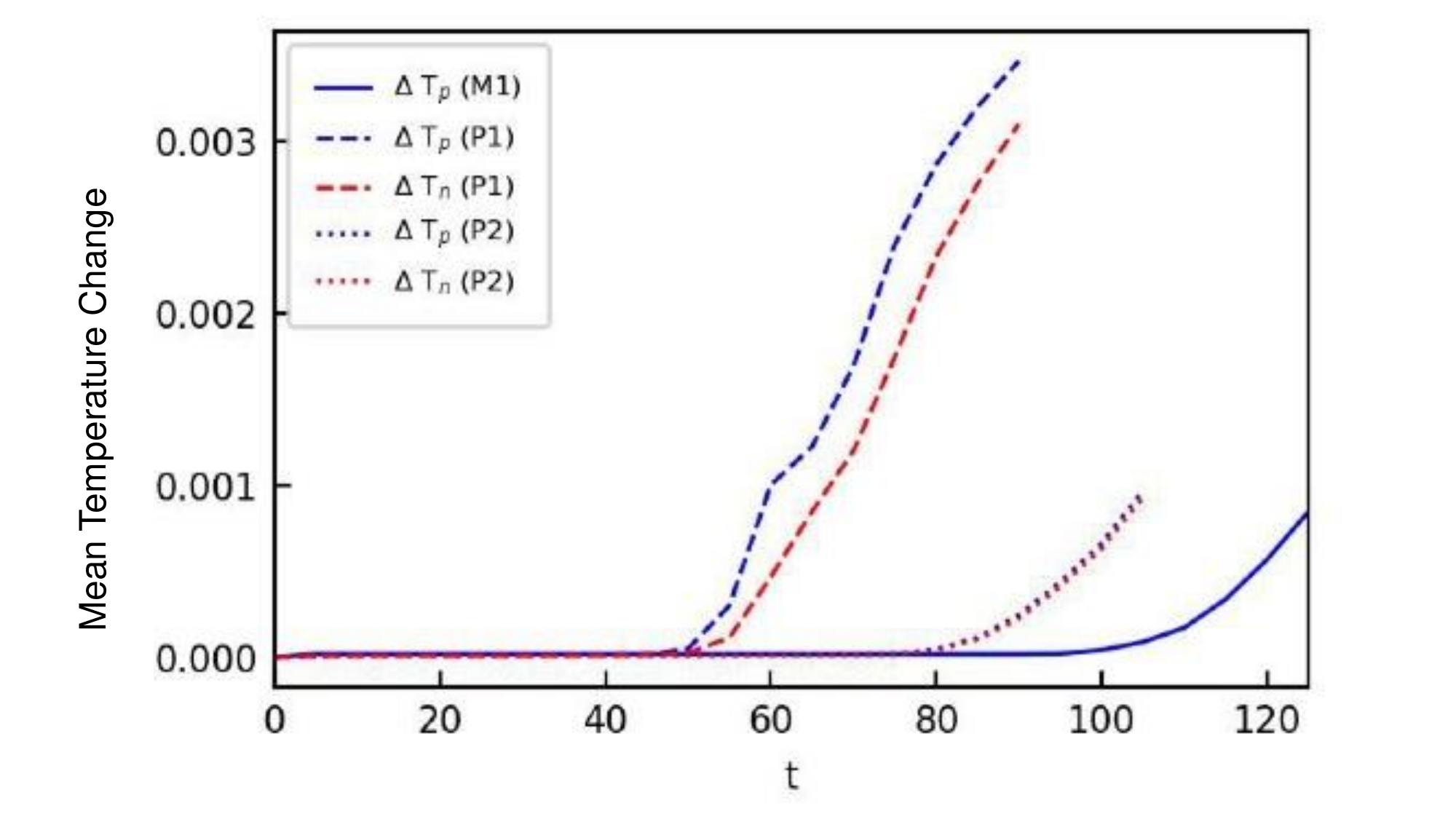}
    \caption{Evolution of the global} mean plasma temperature (blue) and mean neutral temperature (red) of cases M1 (solid line), P1 (dashed line) and P2 (dotted line).
    \label{fig:11}
\end{figure}

Figure \ref{fig:11} shows the changes of the mean plasma and neutral temperatures, averaged over the whole domain, of all three simulations. Due to the normalization based on the Alfvén speed, the initial mean plasma temperature is slightly higher in the MHD case, where $T_p (t = 0) = 0.0833$, than in the PIP cases, where at $t = 0$ $T_p = T_n = 0.0758$. The time variation of the mean temperature is very small, due to averaging across a large portion of the domain where the plasma is unperturbed. Comparing similar stages of the instability, the plasma temperature of M1 increases up to $T_p = 0.0842$ with a rise of 1\% at $t = 120$, and a similar increase is seen for P2, where $T_p = 0.0768$ (1.3\%) at $t = 105$, while for case P1 $T_p = 0.077$ (1.6\%) at $t = 65$. Due to the original difference in background temperature, the mean temperature of case M1 is higher than P1 and P2, but the PIP cases show a temperature increase that is inversely proportional to the two-fluid coupling.

\section{Summary and Discussion}
\label{discussion}

The kink instability of flux ropes may take place in several layers of the solar atmosphere, but has been well modelled only in coronal plasmas. Due to the lack of numerical models, it is still unknown of how this instability develops in a plasma that is only partially-ionised, such as the chromospheric plasma. In this work we present a study of the kink mode developing in PIP, as the growth of the instability is observed in presence of partial ionisation and compared to the evolution in a fully-ionised plasma. Our results can be summarised as follows.
\begin{itemize}
    \item The growth time scale of the helical kink instability increases with the collisional coupling. The nonlinear phase begins earlier in PIP, and is faster for the most decoupled case (P1). This depends on the scaling with respect to the Alfv\'en speed, which is larger the lower the ion fraction and the coupling with neutrals. Consequently, the growth rate is larger the lower the ion-neutral collisional coupling. In all three cases, the estimated growth rate of the instability, normalised by the Alfvén time, matches the values found in \cite{2008A&A...485..837B} ($\sigma \sim 0.05-0.20$) and \cite{2009A&A...506..913H} ($\sigma \sim 0.075$). In the limit of a completely decoupled system ($\alpha_c = 0$), as pointed out by \cite{2021PhPl...28c2901M,2022PhPl...29f2302M}, we expect the instability of the growth rate to be the fastest: this can be considered as a fully-ionised plasma, whose total density is equal to the density of the partially-ionised plasma.
    \item The loss of magnetic energy during reconnection results in a larger increase of kinetic energy for the MHD case and of internal energy for the PIP cases. One of the reasons for the faster temperature increase of the PIP cases is that there are more heating terms, such as frictional heating, contributing.
    \item The maximum Ohmic heating is similar in both PIP cases and locally bigger than frictional heating, but frictional heating is more distributed and its total value in the system is larger. In PIP cases, collisional frictional heating is the major contribution to heating. In the weakly coupled regimes explored in this work, collisional frictional heating is bigger in case P2 than P1, exhibiting an increase that is directly proportional to the collisional coupling. In more strongly coupled systems, collisional frictional heating is expected to reduce, following the increasingly smaller drift velocity between charges and neutral.
\end{itemize}

These results suggest that flux ropes forming in partially-ionised plasmas are more unstable to the kink mode, compared to fully-ionised plasmas of equal total density.
This could lead to a faster release of energy following the shorter time scale of the instability, and contribute to the onset of chromospheric jets, where plasmoid formation has been already observed as a product of reconnection \citep{2011PhPl...18k1210S} and chromospheric micro- and mini-filament eruptions \citep{2015Natur.523..437S,2016ApJ...821..100S,2016ApJ...828L...9S,2019Sci...366..890S,2020ApJ...893L..45S}.
Using the values of temperature and plasma $\beta$ imposed as initial conditions, %($B = 200$ G, $n = 10^{20}$ m$^{-3}$)
the Alfvén speed is $\sim 40$ km s$^{-1}$. The typical length of a mini-filament is $\sim 8 \cdot 10^3$ km \citep{2015Natur.523..437S}, therefore we estimate the initial flux rope radius to be $\sim 400$ km. This leads to a time scale $\tau_A \sim 10$ s. %A mini-filament of fully-ionised plasma (M1) would reach the nonlinear stage of the helical kink instability (i.e. the reconnection onset) in about 15 minutes, against a time frame of $\sim 13$ minutes for a partially-ionised plasma with intermediate coupling (P2) and only 8 minutes for a weakly-coupled plasma (P1). 
For typical chromospheric parameters, including a total number density $n = 10^{20}$ m$^{-3}$, the ion-neutral collisional frequency is calculated as
\begin{equation}
    \nu_{in} = n_n \sqrt{\frac{8 K_B T}{\pi m_{in}}} \sigma_{in},
\end{equation}
where $n_n$ is the neutral number density is a fraction of the total density, $m_{in} = m_i m_n / (m_i + m_n) \sim m_i$ and $\sigma_{in} = 5 \cdot 10^{-19}$ m$^2$ is the cross section, and varies in the interval $10^3 - 10^6$ s$^{-1}$ \citep{2005A&A...442.1091L, 2012ApJ...747...87K}. In non-dimensional form, this would correspond to an $\alpha_c \sim 10^4 - 10^7$, which is a strongly coupled regime. The effects of weak ionisation and coupling are enhanced in much smaller structures: for flux ropes with radius on the order of 10s of meters 
%the ion-neutral collisional rate is of the order of 1 collision s$^{-1}$. T
the dimensional ion-neutral collisional frequency $\nu_{in} = 0.09-0.9$ s$^{-1}$. %for our PIP cases is compatible with flux ropes of radius $\sim 4-40$ m. 
This however has to be considered as a lower limit for which strong two-fluid effects can be observed: both ion fraction and plasma $\beta$ can be orders of magnitude lower than the values imposed here, which would lead to stronger local two-fluid effects. It has to be mentioned that in this work we haven't explored the regime between weakly and strongly coupled plasmas ($\alpha_c = 10^2 - 10^3$), which would lead to the enhancement of partial ionisation effects in larger structures. We defer this to future studies. %Kinking flux ropes in chromospheric current sheets could lead to a faster release of energy following the shorter time scale of the instability, and contribute to the onset of chromospheric jets, where plasmoid formation has been already observed as a product of reconnection \citep{2011PhPl...18k1210S} and chromospheric micro- and mini-filament eruptions \citep{2015Natur.523..437S,2016ApJ...821..100S,2016ApJ...828L...9S,2019Sci...366..890S,2020ApJ...893L..45S}. 

Partial ionisation is also responsible for more heating than what produced during the instability growth in fully-ionised plasmas. The new heating term is generated by the drifts between charges and neutrals (frictional heating) that is dominant over the Ohmic heating. Flux ropes are therefore more heated in partially-ionised plasmas, as more energy is converted into heat than kinetic energy, unlike the fully-ionised coronal processes. This result identifies an important contribution to the heating of the chromosphere, and can be connected to the works of \cite{2005ApJ...618.1020G} and \cite{2007ASPC..368..107H}, whose simulations suggest that most of the heating in the solar atmosphere occurs at chromospheric heights. The localised heating, produced by the small-scales explosive events that are triggered by the instability of chromospheric flux ropes (e.g. chromospheric jets), provides an important contribution to the chromospheric heating, and could also significantly fuel the coronal heating \citep{2009ApJ...701L...1D}.

The study of mini-filaments of finite length would require a different type of boundary conditions at the footpoints (i.e. conductive walls, \citealp{2009A&A...506..913H}, line-tying conditions, \citealp{1979SoPh...64..303H}). While periodic boundaries are still appropriate to examine how partial ionisation changes the evolution from the standard MHD case. This allows us to infer that even with more realistic boundaries we will see faster growth of the instability and more heating. We defer the use of a more realistic setup to future studies.

\section*{Acknowledgements}
The authors thank Prof. Rony Keppens and Dr. Claire Foullon for the precious comments and suggestions that led to the completion of this study. GM is supported by the NASA Grant No. 80HQTR21T0087. AH and BS are supported by STFC Research Grant No. ST/R000891/1 and ST/V000659/1. For the purpose of open access, the authors have applied a ‘Creative Commons Attribution (CC BY) licence to any Author Accepted Manuscript version arising. This work used the DiRAC@Durham facility managed by the Institute for Computational Cosmology on behalf of the STFC DiRAC HPC Facility (www.dirac.ac.uk). The equipment was funded by BEIS capital funding via STFC capital grants ST/P002293/1, ST/R002371/1 and ST/S002502/1, Durham University and STFC operations grant ST/R000832/1. DiRAC is part of the National e-Infrastructure. This research also used resources of the National Energy Research Scientific Computing Center (NERSC), a U.S. Department of Energy Office of Science User Facility located at Lawrence Berkeley National Laboratory, operated under Contract No. DE-AC02-05CH11231.

\section*{Data Availability}

The data that support the findings of this study are available from the corresponding author upon reasonable request. The (P\underline{I}P) code is available at the following url: \url{https://github.com/AstroSnow/PIP}. Details of the code and equations are available in \citet{2016A&A...591A.112H}.

\bibliography{sample631}{}
\bibliographystyle{aasjournal}

\end{document}